\documentclass[aps,prl,showpacs,showkeys,twocolumn]{revtex4-1} 
\usepackage{graphicx,amsmath,hyperref}

\def\ket#1{|\,#1\,\rangle}
\def\bra#1{\langle\, #1\,|}

\def\opone{\leavevmode\hbox{\small1\kern-3.8pt\normalsize1}}

\newcommand{\beq}{\begin{equation}}
\newcommand{\eeq}{\end{equation}}
\newcommand{\ba}{\begin{eqnarray}}
\newcommand{\ea}{\end{eqnarray}}
\newcommand{\bea}{\begin{eqnarray}}
\newcommand{\eea}{\end{eqnarray}}
\newcommand{\bma}{\begin{subequations}}
\newcommand{\ema}{\end{subequations}}
\newcommand{\bwt}{\begin{widetext}}
\newcommand{\ewt}{\end{widetext}}
\def\abs#1{|\,#1\,|}

\begin{document}

\title{Multi-qubit time-bin  quantum RAM}

\author{E. S. Moiseev}
\affiliation{Institute for Quantum Science and Technology, University of Calgary, Canada}
\affiliation{Kazan  (Volga Region) Federal University, Russia}
\author{S. A. Moiseev}

\email[]{samoi@yandex.ru}
\affiliation{Kazan Quantum Center, Kazan National Research Technical University,  Kazan, Russia}

\pacs{ 03.67.Hk, 03.67.Lx, 42.50.Md, 42.50.Ex}
\keywords{quantum information, optical quantum memory, cavity QED, quantum random access memory, photon time-bin qubit, photonic molecular.}

\begin{abstract}
We have proposed a scheme of multi-qubit \textit{quantum random access memory} (qRAM) based on the impedance matched photon echo quantum memory incorporated together with the control three-level atom in two coupled QED cavities.  
A set of matching  conditions for basic  physical parameters of the qRAM scheme that provides  an efficient quantum control of the fast single photon storage and read-out has been found. 
In particular, it was found that the qRAM operation is determined by the properties of the photonic molecular realized in the qRAM dynamics. Herein, the maximal efficiency of the qRAM is achieved  when the cooperativety parameter of the photonic molecular equals to unity that can be easily experimentally implemented. 
The quantum address  of the stored photonic qubits can be put into practice  when the address is encoded in photonic multi-time-bin state.
We discuss the advantages of the qRAM in terms of working with multi-qubit states and the implementation by current quantum technologies in the optical and microwave domains. 
\end{abstract}

\maketitle

\section{Introduction}
The elaboration of the universal quantum computing and quantum communication  requires  qRAM for the implementation of basic protocols such as a quantum search over the classical database, discrete logarithms, quantum Fourier transformation,  collision finding and quantum digital signature \cite{QSearch-Lloyd-PRL-2008, Collision-finding-1997, Log-2007, QSclarke2012}.
The qRAM was proposed by Giovanetti et al. \cite{QRAMPRL2008, *QRAM-Lloyd-PRA-2008} and developed  \cite{qRAM-China} in the bucket-brigade architecture. 
The qRAM provides the arbitary access to the  data cells $\left(\ket{D_n}_d\right)$ by the quantum superposition of the address states ($\ket{\Psi^{a}} =\sum_n\alpha_n \ket {\psi_n^a}$): $ \ket{\Psi^{a}} \xrightarrow{\text{qRAM}} \sum_{n=1}^{N} \alpha_n \ket {\psi_n^a} \ket{D_n}_d$ \cite{QRAMPRL2008, *QRAM-Lloyd-PRA-2008}.   
Exponential speed-up over the classical computation requires an operation of qRAM with  large number of qubits. 
Such qRAM should be also compact and integrable in the quantum network \cite{Kimble2008QI} and in the developing hybrid quantum circuits \cite{Hybrid-RMP-2013}. 
We propose using the photon echo based \textit{quantum memory} (QM) technique \cite{Moiseev2001, Tittel2010} for the implementation of such qRAM.
This choice seems reasonable due to a number of promising experimental results obtained in this QM technique for the effective storage of single photon fields. Herein, the photon echo QM technique has  demonstrated the record efficiency \cite{Hedges2010, Hosseini2011, *Hosseini2011QM, *GEM-Cold-Atom-NJP-2013}, and the optical multimode quantum storage \cite{Riedmatten-PRL2014, Sinclair-PRL-2014} of up to 1000 temporal light modes \cite{Damon2011}.  
More recently the photon echo QM was developed \cite{ SimonPRA2010,*MoiseevPRA2010} and demonstrated \cite{KrollPRL2013} for the atomic ensembles in the impedance matching QED cavity that opens a practical way for implementing compact multi-qubit QM devices.
Such photon echo QM is also applicable for the microwave spectral range [\cite{MicrowaveMemoryPRL2013, *AfzeliusNJP2013,*Gerasimov-PRA2014, MicrowaveMemoryPRX2014}, for the integration in quantum computer schemes \cite{MoiseevJOPB2012} and it can  work efficiently with intensive quantum light fields \cite{Moiseev2004, *KrausPRA2006,*Chaneliere2014}.

In this work we propose the architecture of multi-qubit qRAM consisting of the photon echo QM unit containing $N$ resonant atoms and control three-level atom situated in two coupled QED cavites. 
We found the optimal physical parameters of the qRAM providing the maximal performance and proposed a scheme for addressing storage of the photon qubit. 
We also discussed potential implementations  of the proposed qRAM in the optical and microwave domains on the basis of current technologies.   

\section{Principal scheme} 
The diagram of the proposed qRAM is shown in Fig. \ref{Scheme}.
The scheme operation is controlled by the three-level atom. 
The control atom is strongly coupled to the QED cavity mode that is used for the implementation of the transistor effect for the quantum transport of the input photon to the QM atomic ensemble (there are a number of different physical implementations for a three-level quantum system in the resonant cavity which could be discussed by taking into account  certain physical parameters of the analyzed qRAM).
If the control atom is in the ground state, it will reflect the incoming  photon back, however, if the atom is far off the resonance, the incoming photon will be ideally recorded into the QM unit. 
We show how the controlled photon transport can be made reversible and efficient for the implementation of qRAM.


By assuming that the control atom is in the ground state $\ket{g_c}$, the interaction of the input photon with resonant QED cavities and all atoms is described by the wave function:


\begin{figure}[htdp]
\begin{center}
\includegraphics[width=0.48\textwidth]{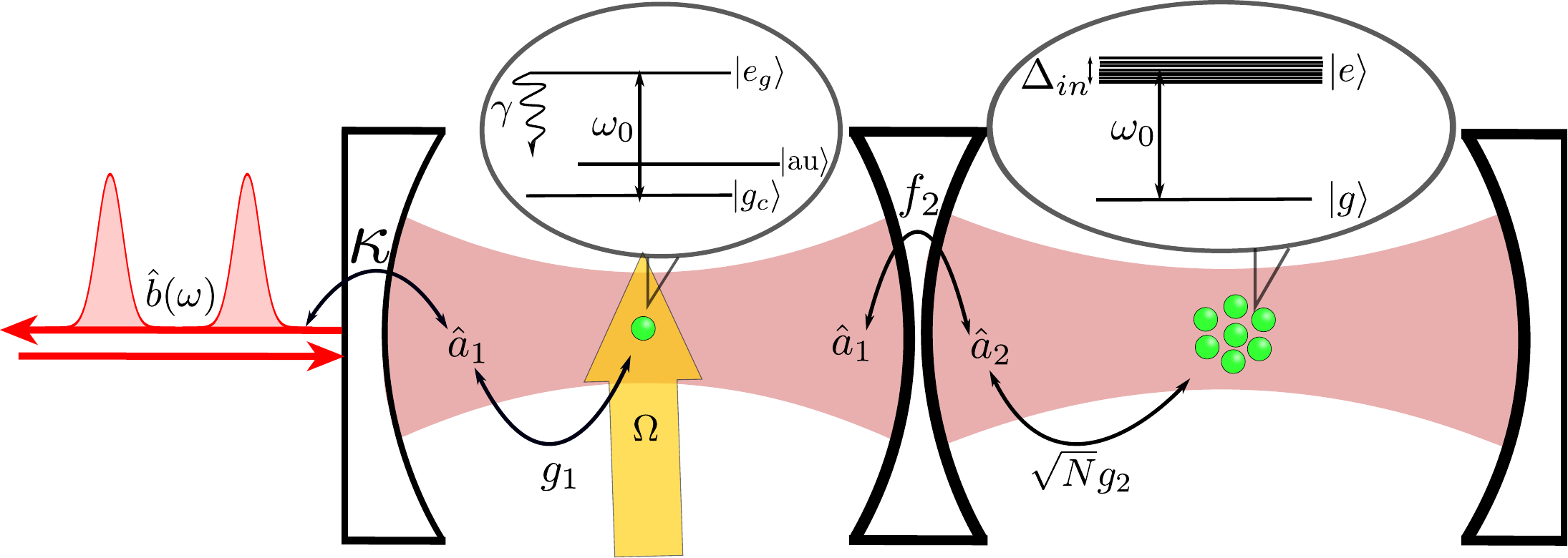}
\end{center}
\caption{(Color online)
Free propagating modes $\hat{b}(\omega)$ are coupled to the  double sided QED cavity with the mode $\hat{a}_1$. This cavity contains a single three-level atom with the quantum transition ($\ket{g_c}\leftrightarrow \ket{e_c}$) resonant to the cavity mode $\hat{a}_1$ characterized by the  frequency $\omega_0$, coupling constant $g_1$, and decay rate $\gamma$ due to the interaction with transverse field modes $\hat c_m(\nu)$; 
$\kappa$ is the decay rate of the cavity mode into the free propagating modes $\hat b (\omega)$, arrow with $\Omega$ means the control laser field with the Rabi frequency on the transition $\ket {g_c} \leftrightarrow \ket {au_c}$.
The other side of the cavity is coupled (with the rate $f_2$) to the  second cavity with a two-level atomic ensemble. The atoms have the inhomogeneous broadening  $\Delta_{in}$ of the resonant line and the collective coupling constant $\sqrt{N}g_{2}$ for the interaction with the cavity mode. }
\label{Scheme}
\end{figure}


\begin{figure}[htdp]
\begin{center}
\includegraphics[width=0.48\textwidth]{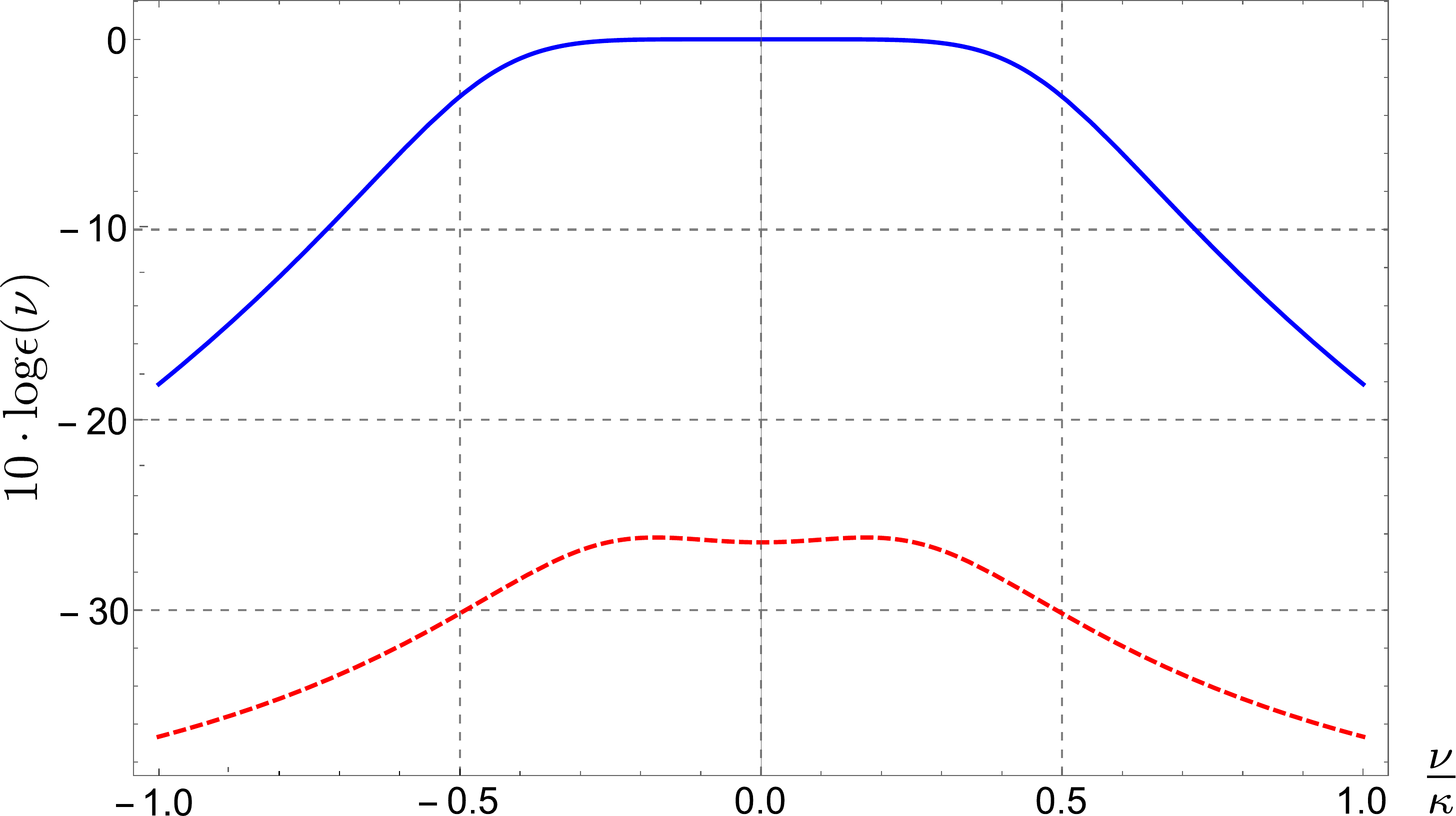}
\end{center}
\caption{(Color online) Spectral transfer function in the logarithmic scale $10\log\epsilon(\nu/\kappa)$ for the single photon storage (blue solid line)  and for the blockade (red dashed line). The wide spectral window for the photonic storage is implementing  by three impedance matching conditions: $C_{QT}=1$ and Eqs. (7),(8). The strong blockade of the memory process is represented for the  single atom cooperativity factor $C=10$ and $\gamma=\kappa$.}
\label{Z-function Blockade}
\end{figure} 

\bea{}
\ket{\Psi(t)}=\bigg(\beta_c  \hat S_+^0+\sum_{j=1}^{N}\beta_{j}\hat S_{+}^j+
\alpha_1 \hat a^{\dagger}_1 
+\alpha_2 \hat a^{\dagger}_2\nonumber \\
+\int d\nu \alpha_{\nu} \hat b^{\dagger}(\nu) + \sum_m \int d\omega r^{m}_{\nu}\hat{c}^{\dagger}_{m}(\nu) \bigg) \ket{0}_f\ket{g_c}\ket{g},
\label{WF}
\eea{}

\noindent
with the normalization  
$\abs{\beta_c}^2 + \sum_{j=1}^{N}\abs{\beta_j}^2 + \abs{\alpha_1}^2+
\abs{\alpha_2}^2+ \int d\nu \abs{\alpha_\nu}^2 + \sum_{m}\int d\nu  \abs{r^{m}_{\nu}}^2 =1$ and the initial quantum state: 
$\ket{\Psi(t\rightarrow-\infty)}=\ket{\psi_{in}(t)}_f=\int d\nu\alpha_{\nu}^0(t)\hat{b}_{\dagger}(\nu)\ket{0}_f$;
$\alpha_{\nu}^0=\alpha_\nu(-\infty)$ characterizes the state of the input photon wave packet;$\int d\nu \abs{\alpha_{\nu}^0}^2=1$;
$\beta_c = \beta_j=\alpha_{1,2}=r^{m}_{\nu}=0$;  
$\ket{0}_f$ is the vacuum state of the input light modes; $\ket{g_{c}}$ and $\ket{g}\equiv \prod_{j=1}^N \ket{g_j} $ are the ground states of the control atom-c and QM atoms,
where $\hat{S}_z^0$, 
$\hat{S}_z^j$ are z-projections of the effective spin 1/2 operators on the $\ket{g_{c,j}} \rightarrow \ket{e_{c,j}}$ transition,
$\hat{S}^j_+$, $\hat{S}^j_-$ and  $\hat{S}^0_+$, $\hat{S}^0_-$  are the transition spin operators of $j$-th and control atoms;
$\hat{a}^{\dagger}_{1,2}$ and $\hat{a}_{1,2}$ are arising and decreasing operators of the $1$-st and $2$-nd cavity field modes;
$\hat b^{\dagger}(\nu),\hat b(\nu)$ are the bosonic operators of free propagating modes ($\left[ \hat b(\nu),\hat b^{\dagger}(\nu')\right]= \delta (\nu-\nu')$); $\hat{c}_{m}(\nu),\hat{c}^{\dagger}_{m}(\nu)$ are the bosonic operators of the bath interacting with the control atom in the first cavity ($\left[ \hat{c}_{m}(\nu),\hat{c}_{n}^{\dagger}(\nu')\right]= \delta_{m,n}\delta (\nu-\nu')$). 

By using the well-known input-output formalism \cite{walls2007quantum}, we solve the Schr\"odinger equation 
for the wave function $\ket{\Psi(t)}$  that gives the following amplitudes for the large time of the interaction $t >\delta t$ (where $\delta t$ is the temporal duration of the input photon wavepacket):

\noindent
\beq{}
\beta_j(\Delta_j,t) = i\sqrt{2\pi\kappa} \frac{g_2}{f_2}  F(\Delta_j) \alpha^0_{\Delta_j} e^{-i(\Delta_j-i/T_2) t},
\eeq{}
 
\beq{}
F(\Delta)=\frac{f^{2}_2}{(Ng^{2}_2 \tilde{G}(\Delta) -i\Delta)( \frac{\kappa}{2} +   \frac{i g^{2}_1}{ \Delta-\delta+\frac{i}{2}\gamma }-i\Delta) + f^{2}_2 }, 
\eeq{}

\noindent
where $\tilde{G}(\Delta)=\int d \nu \frac{G(\nu)}{\epsilon+i(\nu-\Delta)}$, $G(\nu)$ is the formfactor of the \textit{inhomogeneously broadened} (IB) resonant line;
we also used the continuous limit for the large number of atoms in QM: $\beta_j(\Delta_j,\tau) \rightarrow \beta({\Delta},\tau)$, $\sum^{N}_{j=1}... \rightarrow N \int d\Delta G(\Delta)...$, $T_2$ is the decoherence time of the atomic ensemble in QM.

We find the probability of the photon transfer in QM using $\beta(\Delta,\tau)$:
$P_a (t>\delta t)=\sum_{j=1}^{N}
|\beta_j (t)|^2 = \int_{-\infty}^{\infty} d\Delta \epsilon(\Delta)|\alpha^{0}_{\Delta} |^2$. 
Analogous to any classical spectral device,   qRAM can be characterized by its spectral efficiency   
$\epsilon(\Delta)= 2\pi N \kappa \abs{\frac{g_2}{f_2}}^2 G(\Delta)\abs{F(\Delta)}^2$ for the single photon storage. 
For simplicity but without loss of generalization, we assume IB to be Lorentzian $G(\nu)\equiv G_L(\nu)=\frac{\Delta_{in}}{\pi \left( \nu^2 + \Delta_{in}^2\right)}$  with the  bandwidth $\Delta_{in}$.
For the narrow bandwidth of the input photon field  ($\delta\omega_f\sim \delta t^{-1}\ll\kappa,\Delta_{in}$), we find the following spectral efficiency  

\bea{}
\epsilon(\Delta \approx 0)
=\frac{4 C_{pm}}{(1 +  C_{pm} +\frac{\gamma^2 C}{\delta^2+(\gamma/2)^2})^2+ (\frac{2\delta\gamma C}{ \delta^2+(\gamma/2)^2})^2 }, 
\label{SpectralFunc}
\eea{}

\noindent
determining the main properties of qRAM,  
where  $C =\frac{g_{1}^{2}}{\kappa \gamma}$ is a well-known single atom cooperativity factor; we also introduced the \textit{ cooperativity factor of photonic molecular} for the analyzed qRAM:
$C_{pm} = \abs{f_{2}}^2/( \kappa\frac{Ng_{2}^2}{2 \Delta_{in}})$.
The factor $C_{pm}$ reflects the quantum  and dissipative properties of the photon in two coupled QED cavities, since $f_2$  is the interaction constant between the two quantum states of the photon,  i.e. the coupling constant between the two states of the photonic molecular (the photon exists in the first or in the second QED cavity), while $\kappa$ and $N g_2^2/(2\Delta_{in})$  are the decay constants of the photon states in these QED cavities.

By analyzing Eq. (\ref{SpectralFunc}), we find two basic regimes of the qRAM operation: 1) Storage of a single photon in the  QM ensemble and 2) Blockade of the photon storage (retrieval) in (from) the QM ensemble. 


1) Perfect storage is implemented for the large spectral detuning of the control atom
$\abs{\delta}>>\gamma C$.  It can be implemented by  transferring the atom in the third ancillary state $\ket{au_c}$ (see Fig.1) (\ref{SpectralFunc}) where the resonant  storage efficiency is: 
\bea{}
\epsilon(0)\mid_{|\delta|>>\gamma C}\equiv \epsilon_T(0)= 4 C_{pm}/(1 +  C_{pm})^2,
\label{efficiency}
\eea{}
\noindent
that shows the ideal transmission (index $"T"$ in $\epsilon$) of single photon in the QM ensemble (\ref{efficiency}) at  $C_{pm}=1$, which is the first \textit{impedance matching} (IM) condition of the effective qRAM. 
Equation (\ref{efficiency}) with $\epsilon(0)=1$ at $C_{pm}=1$ reproduces  properties, which are similar to the properties of the impedance matching photon echo QM \cite{SimonPRA2010, MoiseevPRA2010}  but in the presence of the additional strongly interacting QED cavity mode.  
Herein, the impedance photon echo QM \cite{SimonPRA2010, MoiseevPRA2010} is characterized by the different coupling constant $2 N \abs{g_2}^2$ and the following two decay constants $\kappa$, $\Delta_{in}$. 
These parameters satisfy the impedance matching condition  $\kappa\Delta_{in}=2 N \abs{g_2}^2$
(it is possible to generalize this condition and facilitate its implementation by transferring to the off-resonant Raman echo QM scheme \cite{MoiseevPRA2013}).  
The impedance matching condition of qRAM couples somewhat diffent physical parameters and we will see how this new matching condition  provides  the convenient implementation of qRAM operations. 
 
Analyzing $\epsilon(\nu)$ for the transfer process we found the second 
 

\bea{}
\frac{N\abs{g_{2}}^{2}}{\Delta_{in}} = \frac{\Delta_{in} \kappa/2}{(\Delta_{in}+ \kappa/2)},
\eea{}

\noindent
and third  

\bea{}
\Delta_{in} = \kappa/2,
\eea{}
\noindent
IM conditions, where the spectral quantum  efficiency $\epsilon(\nu)$  has the almost ideal flat spectral behavior  (see Fig \ref{Z-function Blockade}) around $\nu\approx 0 $ that is given by the expression 

\bea{}
\epsilon_T(\nu) = \frac{1}{1+(\frac{\nu}{\kappa/2})^6},
\eea{}

\noindent
and providing the efficient transfer of the broadband input single photon field. This process transfers the  wave function (1) into the purely atomic state  
$\ket{\Psi(t>\delta t)}\cong\ket{0}_f\ket{au}_c\left( \sum_{j=1}^{N}\beta_{j}(t)\hat S_{+}^j \right) \ket{g}$.


2). In the case of the resonant interaction with the control atom
(when the atom stays in the state $\ket g_c$ where $\delta=0$) we find

\bea{}
\epsilon(0)_{\mid_{\delta=0}} \equiv \epsilon_{B}(0)=
4 C_{pm}/(1 +  C_{pm} +4 C)^2,
\eea{}
that leads to 
\bea{}
\epsilon_{B}(0)_{\mid_{C_{pm}=1}}=(1 + 2 C)^{-2}.
\eea{}
By using  the single atom cooperativity factor $C$ from the experimental data \cite{Hybrid-RMP-2013} for the single atom in the Fabry-Perot cavity $C\equiv C_{opt}=30$ and for the superconducting qubit in the moderate microwave resonator  $C\equiv C_{\mu w}=300$, we get quite good blockade with $\epsilon_{B,opt}=2.6 \cdot 10^{-4}$ and $\epsilon_{B,\mu w}\approx 3 \cdot 10^{-6}$. 
It is important to note that the choice of the most acceptable three-level quantum systems and the type of the QED cavities is still an open question for the further studies, where using photonic waveguides, nano-fibers and surface plasmon polaritons could be also promising for the experimental implementation.   


The most important benchmark of the photon blockade is the reflection of the input photon field. 
The direct usage of the input-output relationship $\alpha_{in}(\nu) + \alpha_{out}(\nu) = \sqrt{\kappa}\alpha_1(\nu)$ and the relation $\alpha_1(\nu)=A_{1,in}(\nu)\alpha_{in}(\nu) $ (see $A_{1,in}(\nu)$ in the supplement material) leads to
\bea{}
\alpha_{out}(\nu)=f_{Bl}(\nu) \alpha_{in}(\nu),
\eea{}
where
\bea{}
f_{Bl}(\nu)=
\frac{i\kappa}{\nu+ \frac{i\kappa}{2} - \frac{g^{2}_1}{ \nu-\delta+i\gamma/2 } - \frac{f^{2}_2}{\nu +i Ng^{2}_2 \tilde{G}(\nu) }} -1 .
\eea{}
Here for sufficiently narrow spectral width of the storage light and resonant interaction with control atoms (${\delta=0}$) we obtain
\bea{}
f_{Bl}(|\nu|<\kappa)\mid_{\delta=0}   \rightarrow 
\left( \frac{1}{\frac{1+C_{pm}}{2}  +2C} -1\right).
\eea{}

Taking into account here the  first matching condition $C_{pm}=1$ we find for Eq.(11)

\bea{}
\alpha_{out} = - \frac{2C}{(1+2C)} \alpha_{in}.
\eea{}

\noindent
The reflection (14) is valid with high accuracy within the spectral width which is comparable with the memory storage window $\sim \kappa$ if all three IM conditions  hold. 
Inserting the same values for the single atom cooperativity factor we obtain  $\abs{\frac{\alpha_{out}}{\alpha_{in}}}^2=0.983$ and $0.996$ for  $C\equiv C_{opt}=30$ and $C \equiv C_{\mu w}=300$, respectively, i.e. the rather strong blockade (see Fig. 2). 

By implementing the CRIB procedure for the retrieval of the stored single photon field, we invert the atomic detuning $\Delta_j \rightarrow - \Delta_j$ at the time $t=\tau$ (in general, one can also use some other experimental methods for rephasing the atomic coherence, such as AFC- or silent echo protocols, but CRIB is easier to be demonstrated due to its perfect time reversibility). 
Initially all free propagation modes and two cavity modes are in the vacuum state and the control atom is in the ground state, while the QM atomic coherence is given by  $\beta_j (\Delta_j,\tau)$. 
Calculating the echo field emission (see the readout stage in the supplement materials), which occurs in the time reversal manner,  we find  the wave function  of the light-atomic system for the time $t\gg 2\tau$ with the following spectral photonic amplitude $\alpha_{\nu}(t)$ is: 

\bea{}
\alpha_{\nu}(t)&=&
-2\pi \kappa N\abs{\frac{g_2}{f_2}}^2 G_L(\nu) F_S(-\nu)F_R(\nu)\nonumber\\
&\alpha &_{-\nu}^0 e^{-i\nu(t-2\tau)-2\tau/T_2},
\label{echo}
\eea{}
\noindent
where indices $S,R$ denote the \textit{storage} and the \textit{retrieval} stages. 
Herein, if the control atom stays in the state $\ket {au}_c$ (i.e., $S,R \rightarrow T$), we find from Eq.(\ref{echo}) the following quantum efficiency of the stored photon retrieval: 

\bea{}
P_{echo \mid_{\delta\omega_f\leq 0.2 \kappa}} \cong \frac{16C_{pm}^2e^{-4\tau/T_2}}{(1+C_{pm})^4}\mid_{_{C_{pm}=1}}
=e^{-4\tau/T_2}, 
\eea{}


\noindent
where 
the wave function of the iradiated photon field is  $\alpha_{\nu}(t)\cong -\alpha_{-\nu}^0\exp\{-i\nu(t-2\tau)-2\tau/T_2\}$, that means the perfect time-reversal retrieval of the initial single photon state under the condition of the long-lived coherence time $2\tau/T_2 \ll 1$. 
The probability of the echo pulse retrieval for different decoherent time $T_2$ and pulse duration $\delta t$  is presented in Fig. 3, where one can find the necessary relation between $T_2$ and other parameters.

\begin{figure}[htdp]
\begin{center}
\includegraphics[width=0.48\textwidth]{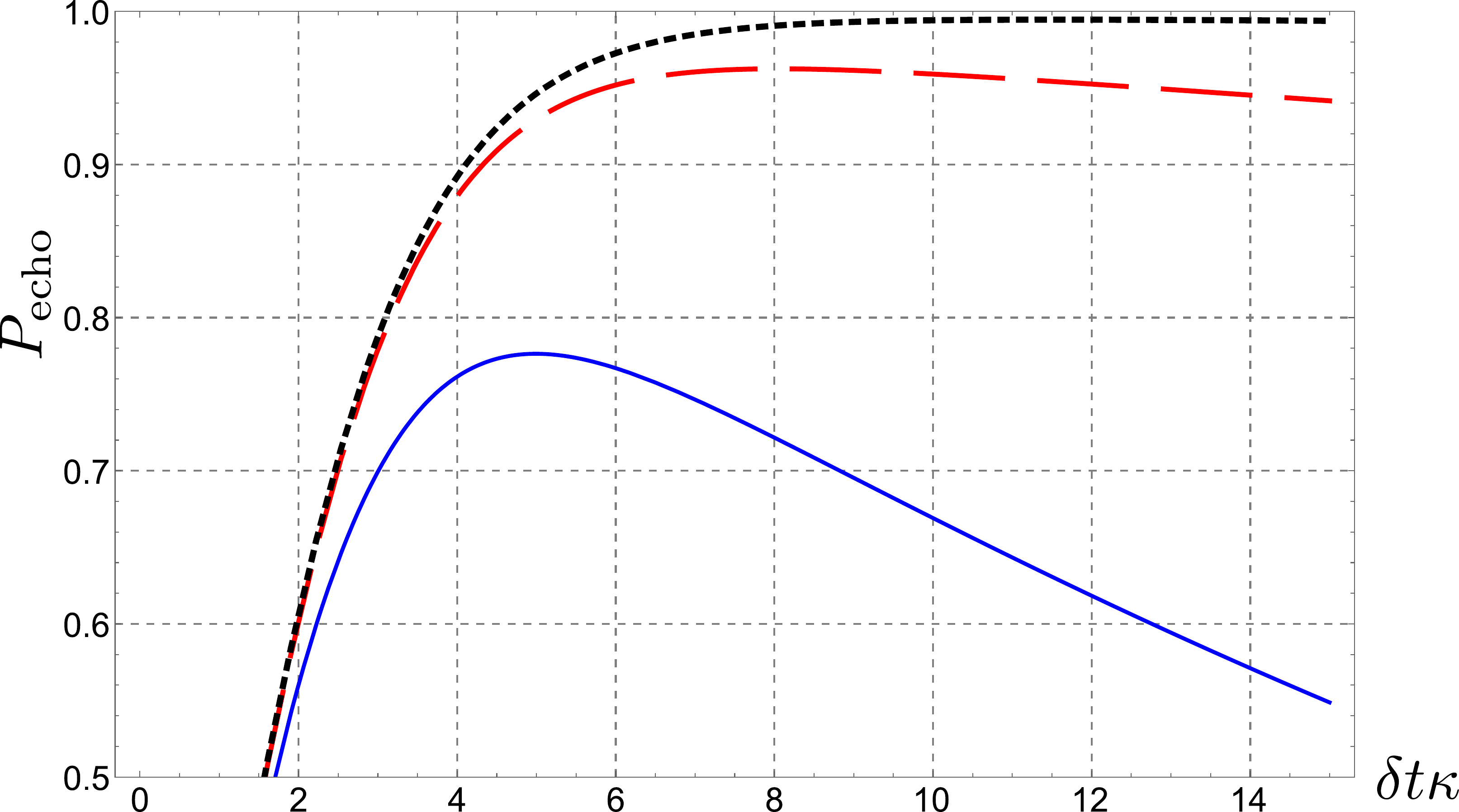}
\end{center}
\caption{(Color online). Probability of  the echo pulse retrieval  as a function of the Gaussian pulse duration $\delta t$ (in units  $\delta t\kappa$) for different atomic relaxation times   $T_2$:  $T_2 \kappa = 10^2$ (blue, solid line); $T_2 \kappa = 10^3$  (red, long dashed line);  $T_2 \kappa = 10^4$ (black, short dashed line)  when three IM conditions hold.}
\end{figure}

For the case, when the control atom is prepared in the ground state $\ket {g_c}$,  where $\delta=0$,  we  implement the  blockade for the photon retrieval (where the index $R\rightarrow B$). Here, under the condition of the perfect storage ($S\rightarrow T$) we obtain for the photon emission probability:

\bea{}
P_{echo} &=&e^{-4\tau/T_2} \int_{-\infty}^{\infty} d\Delta \epsilon_T(\Delta)\epsilon_B(\Delta)|\alpha^{o}_{\Delta} |^2_{\mid_{\delta\omega_f\leq 0.2 \kappa},C_{pf}=1}\nonumber \\
&\cong& \frac{e^{-4\tau/T_2}}{(1+2C)^2}_{\mid_{C\geq 10}}<0.0023, 
\label{echo_probability}
\eea{}
\noindent
that means the strong blockade of the storaged state, when the echo photon is efficiently reabsorbed by the QM atomic ensemble.  
The atomic amplitudes grasp the additional $\pi$-shift : $\beta (\Delta, \tau)e^{i\Delta t}_{t<2\tau}\rightarrow -\beta (\Delta,\tau)e^{i\Delta t}e^{-(t-\tau)/T_2}_{t>2\tau}$ as a result of the echo photon reabsorption.
One can obtain Eqs. (\ref{echo}),(\ref{echo_probability})  in the limit $1/T_2 \rightarrow 0$ by using the general approach based on the time reverse symmetry of light-atoms equations 
describing the qRAM operation for the retrieval stage.


\section{Quantum addressing} Here we demonstrate the quantum addressing for the retrieval stage of the qRAM operation by assuming the rather large  time $T_2$ of the atomic coherence ($T_2\kappa\geq 10^4$, see Fig.3).
\begin{figure}[htdp]
\begin{center}
\includegraphics[width=0.48\textwidth]{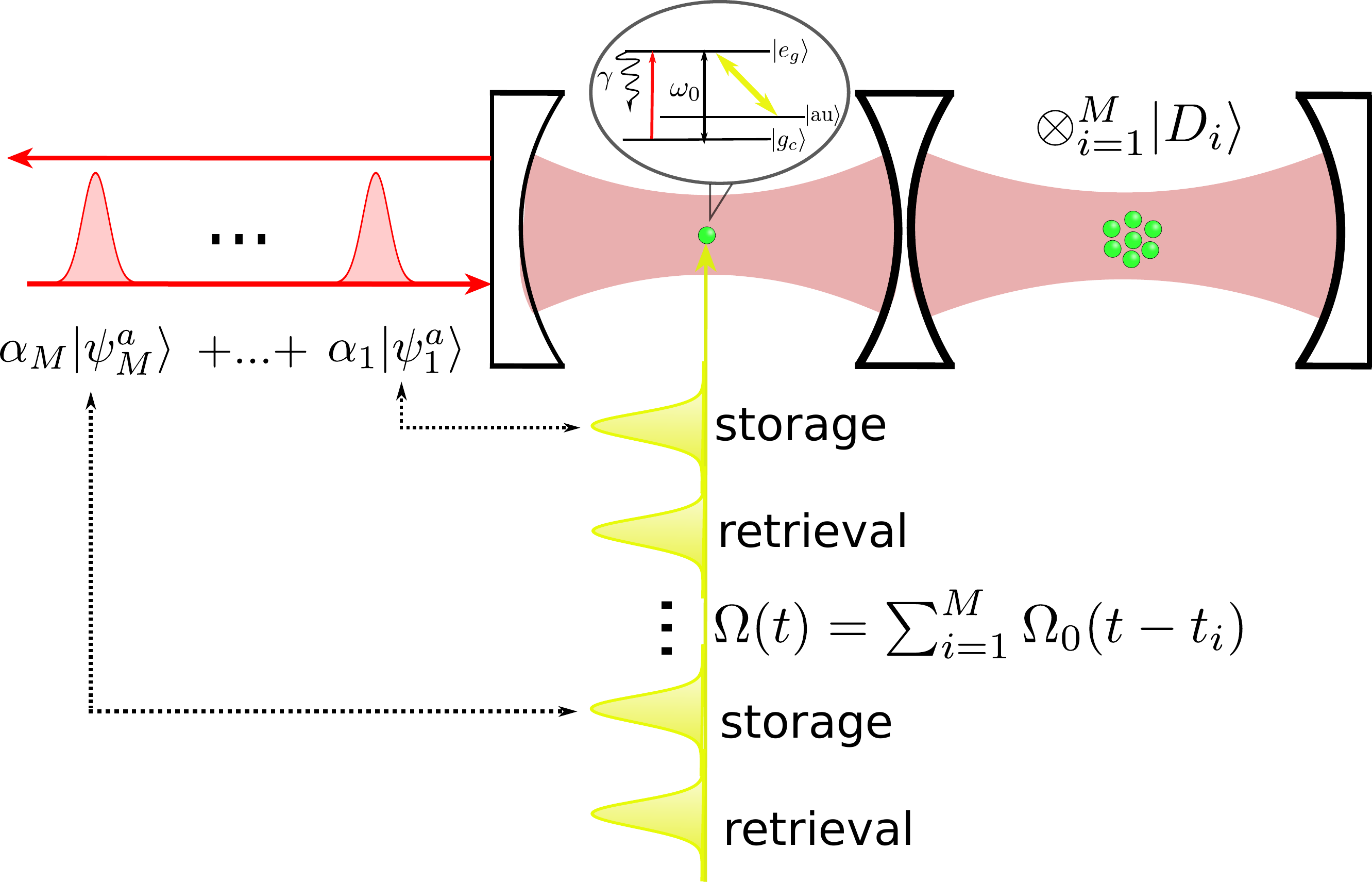}
\end{center}
\caption{(Color online) Quantum addressing is implemented by the synchronization between the readout from the QM unit and the quantum control of three-level atom. 
The quantum information is retrieved  from QM as the sequence of $M$ time-bin single photon. During the readout of each photon the control atom is transferred into the superposition of states $\ket{g}_c$ and $\ket{\text{au}}_c$ via lambda-transition by the classical control field $\Omega(t)$ and the single photon ($\sum_{j}^{M}\ket{\Psi^{a}_{j}}$) distributed in $M$ time-bins (quantum address). Retrieval pulses return the atom to the ground state before each new addressing wave packet.
}
\label{Scheme-addr}
\end{figure}
Let us assume that $M$ photon qubits prepared in the state $\prod_{n=1}^M\ket{\psi_{in,m}(t-t_m)}_f$ have been stored one by one in the atomic ensemble. Here, the qRAM is excited in the state  
\bea{}
\ket{Q_{RAM}}=\hat {S}^+_{(M)}  (t-t_M)...\hat {S}^+_{(1)} (t-t_1) \ket{g}_c\ket{g}, 
\label{qram_state}
\eea{}
where
$\hat{S}^+_{(m)} (t-t_m)=\sum_{j=1}^N \beta_j^{(m)}(\Delta_j, t-t_m)\hat{S}^j_+ $ describes a collective single atomic excitation caused by the absorption of the $m$-th photonic qubit at the time $t\approx t_m$. 
All M atomic excitations are decoupled from each other for $M\ll N$ and in the case of strong reciprocal  dephasing. 
Such separable quantum  state of M qubits in QRAM can be formally  written as a product
$\ket{Q_{RAM}}=\ket{g_c}\prod_{m=1}^M\ket{D_{m}}$, where $M$ states are orthogonal  $\bra{D_{m'}}\ket{D_{m}}\sim\delta_{m,m'}$.
The state $\ket{D_m}$ can be interpreted as the state of  $m-th$ QM cell where  $t_m$ is a time-label of this cell (rephasing/dephasing of the atomic coherence in this cell).

In the case when the $m'$-th atomic excitation will irradiate the stored photon qubit, when the initial quantum state of QRAM and light is transformed as follows:
\bea{}
\ket{0}_f\ket{Q_{RAM}}&\rightarrow&
-\ket{\psi_{in,m'}}_f \hat{S}^+_{(M)}  (t-t_M)\nonumber\\
&...&\hat{S}^+_{(m'+1)}  (t-t_{m'+1}) \hat{S}^+_{(m'-1)}  (t-t_{m'-1})\nonumber\\
&...&\hat {S}^+_{(1)} (t-t_1) \ket{g_c}\ket{g}
\nonumber\\
&\equiv& -\ket{\psi_{in,m'}}_f\ket{g_c}\ket{\emptyset_{m'}}
\prod_{m\neq m'}^M\ket{D_{m}},
\eea{}
where M-1 photon qubits are stored in the QM and $m'$-qubit has been irradiated in the quantum state $\ket{\psi_{in,m'}}_f$, $\ket{\emptyset_{m'}}$ means the empty $m'$ cell of QM.
Analogously, if $m'$-th and $m"$-th stored qubits are irradiated from QRAM, the output state will be $\ket{0}_f\ket{Q_{RAM}}\rightarrow \ket{\psi_{in,m'}}_f\ket{\psi_{in,m"}}_f\ket{g_c}\ket{\emptyset_{m'}}\ket{\emptyset_{m"}}\prod_{m\neq m',m"}^M\ket{D_{m}}$ with two retrieved photon qubits etc.  

For the addressing qubit we use a single photon wavepacket distributed in $M$ time-bins (M coincides with the number of photon qubits stored in QM) that is described by the following quantum superposition 
$\ket{\Psi^{a}}_f = \sum^{M}_{n=1} \alpha_n \ket{\psi_n^a(t)}_f$   (where $\ket{\psi_n^a(t)}_f=\ket{\psi^a[-(t-(n-1)\tau_o-t_c)]}_f$ and the duration of each time-bin is much smaller than the distance between two nearest time-bins $\delta z/c \ll T_0$, $\sum_{n} \abs{\alpha_n}^2 =1$; 
$\ket{\psi^a(t)}_f = \int dz g(t-z/c,\delta z_f) e^{-i\omega(t-z/c)} \hat{a}^{\dagger}(z) \ket{0}_f $
which is stored within the independent temporal  mode (where  $\hat{a}^{\dagger}(z)$ is the Fourier transform of the creation operator in the momentum space 
$\hat{a}^{\dagger}(z)=\frac{1}{\sqrt{2\pi}}\int dk e^{-i(k-\omega_0/c)z} \hat{a}^{\dagger}(k) $; 
$[ \hat{a}(z), \hat{a}^{\dagger}(z') ]=\delta (z-z'))$; 
$g(t)$ describes the temporal shape of the photon wave packet; $\delta z$ is its spatial longitudinal size). 
At the time $t\approx t_c$ we map the first wave packet on the control atom state $\ket {au_c}$ via Raman transition as it is sketched in Fig. \ref{Scheme-addr}. 
These M photon wave packets provide the orthogonality for all M quantum  addresses. 
Such single photon state can be prepared by using the photon echo QM \cite{MultiPulse-PRA-2004} or by the stimulated rapid adiabatic passage that was successfully implemented with the high efficiency  \cite{multi-time-NJP-2013}.



By taking into account the initial state $\ket {\Psi_{in}} =\ket{\Psi^{a}}_f\ket{Q_{RAM}}$, we consider how the addressing protocol works for the readout process.
At first we analyze the retrieval stage for the $1$-st addressing wave packet ($n=1$: $1\leq n \leq M$) of   $\ket{\Psi^{a}}_f$.  
The first wave packet and the control laser pulse provide all together the Raman resonant transfer of the control atom $\ket {g_c} \rightarrow \ket {au_c}$ that leads to the following state

\bea{}
\ket{\Psi_{in}} &\rightarrow & \ket{\Psi_{1}}
\nonumber \\
&=& \ket{0}_{out,f}\prod_{m=1}^M\ket{D_{m}} \bigg(\ket{g_c}\sum^{M}_{n=2} \alpha_n \ket{\psi_n^a}_f
\nonumber\\ 
&-&\alpha_1\ket{au_c}\ket{0}_f\bigg), 
\label{adressing}
\eea{}

\noindent 
where $\ket{0}_{out,f}$ denotes the vacuum state of the output light field modes. Herein, the field mode of the second  QED cavity is shifted out from the resonance with the field mode of the first QED cavity during the interaction with the control atom that can be implemented by using the acoustic-optical modulator in the second QED cavity.
In particular, the highly efficient mapping (\ref{adressing}) of the photon wave packet on the control atom state $\ket{au_c}$ is possible for the exponentially rising shape $g(-t)$ \cite{OptimalStoragePRL2007}.
The exponential shape $g(t)$ corresponds to the typical  photon irradiation by the single two-level system. 
At the same time $g(-t)$ can be put into practice via various generalized time-reversal CRIB schemes \cite{MoiseevNJP2013, *CampbellPRL2014}.

In the next step we retrieve the first photon qubit from QM by rephasing the atomic coherence in the QM ensemble.  
Here, we return the atomic ensemble in resonance with the two QED cavity modes and rephase only the $1$-st atomic coherence of the state $\ket{\phi_1 (t)}$ during the readout process by inverting the frequency detuning of each atom $\Delta_j\rightarrow-\Delta_j$. 
The atomic rephasing will lead to the transfer of the $1$-st atomic state $\ket{D_1}$  into the free propagating photon wave packet or returns to the atomic ensemble in the case of the atomic blockade. 
These two alternatives of the qRAM operation happen in accordance with the analyzed two basic regimes ($1$-st or $2$-nd) determined by two control atom states ($\ket{g_c}$ or  $\ket{au_c}$).
Eventually these two quantum alternatives lead to the following transformation of the state $\ket {\Psi_{1}}$:

\bea{}
\ket {\Psi_{1}} &\rightarrow& \ket {\Psi_{2}}
=\ket{0}_{out,f}\prod_{m=2}^M\ket{D_{in,m}}
\nonumber\\
 &\{&\alpha_1\ket{au_c}\ket{\emptyset_{1}}\ket{\psi_{in,1}}_f \nonumber\\
&+&\ket{g_c}\ket{\phi_{in,1}}\sum^{M}_{n=2} \alpha_n \ket{\psi_n^a}_f
\}. 
\eea{}

Next we return  the excited state $\ket{au_c}$ of the control atom to the initial state $\ket{g_c}$  by the retrieval of the first time-bin photon wave packet via applying an additional laser pulse on the transition $\ket{au_c}\rightarrow\ket{e_c}$ (see Fig. \ref{Scheme-addr}). This atomic transfer $\ket{e_c}$ leads to the following transition

\bea{}
\ket {\Psi_{2}} &\rightarrow& \ket {\Psi_{3}} 
\nonumber \\
&=&  \prod_{m=2}^M\ket{\phi_{in,m}} \{-
\alpha_1\ket{g_c}\ket{\emptyset_{1}}\ket{\psi_1^a}_f\ket{\psi_{in,1}}_f
\nonumber\\
&+&\ket{g_c}\ket{\phi_{in,1}}\ket{0}_{out,f}\sum^{M}_{n=2} \alpha_n \ket{\psi_n^a}_f
\}. 
\label{qRAM_operation}
\eea{}

The light field component of the state (\ref{qRAM_operation}) is characterized by the two entangled photon wave packets irradiated by the control  atom and by the atomic ensemble of qRAM at two different moments of time.      
We repeat this process one by one for the second and all others  $M-1$ time-bin addressing wave packets that leads to the following output state:

\bea{}
\ket {\Psi_{in}} &\rightarrow&  \ket {\Psi_{out}}  = 
\nonumber\\ 
&-&  \sum^{M}_{n}\alpha_n\ket{g_c}\ket{\emptyset_{n}}\ket{\psi_n^a}_f
\ket{\psi_{in,n}}_f\prod_{m\neq n}^M\ket{D_{m}}.
\label{finalstate}
\eea{}

As it is seen in Eq.(\ref{finalstate}), 
the two photon field is irradiated in the entangled state of the addressed and retrieved photons. 
The final state is a result of the pure unitary
evolution leading to the quantum superposition of M two-photon states - $\ket{\psi_n^a}_f$,  $\ket{\psi_{in,n}}_f$ with amplitudes determined by the addressing states that accomplishes the qRAM operation on the multi-qubit QM. 
We note that other $(M-1)$-stored qubits $\ket{\phi_{in,m}}$ become  also entangled with the two photon qubits that is done through the superposition of the states $\ket{\psi_n^a}_f\ket{\psi_{in,n}}_f\ket{\phi_{in,1}},...,\ket{\phi_{in,n-1}},\ket{\phi_{in,n+1}},\ket{\phi_{in,M}}$. 
It is obvious, that the performed analysis is valid for the arbitrary initial quantum state of the QM atomic ensemble, for example, entangled with another freedom degrees of its environment. Finally, we note that by exploiting the time-reverse symmetry of the light-atom interaction we can  implement the quantum addressing storage of the photonic qubit. 
Thus we showed the possibility of the effective implementation of the multi-qubit qRAM based on the photon echo QM and three-level atom incorporated in two coupled QED-cavities.

\section{Experimental issues}
Nowadays it seems promising to implement the proposed qRAM  in the circuit and nano-optical QED schemes. 
All basic required hardware elements are implemented in the circuit QED: system of two coupled cavities with quantum emitters \cite{Coupled-QED-Cavities-PhysRevX-2014}, the multi-qubit echo memory of the spin ensemble coupled to the cavity is under the active development \cite{MicrowaveMemoryPRL2013, *AfzeliusNJP2013,*Gerasimov-PRA2014, *MicrowaveMemoryPRX2014}  as well as the coherent transfer of the single artificial atom population \cite{johnson2010quantum, EIT-CurcuitQED-PRL-2010, AT-CurcuitQED-PRL-2009}.  
The technique of the multi-time-bin photon generation,  which  was successfully demonstrated in the cavity QED \cite{multi-time-NJP-2013}, can also be applied to microwave domain \cite{johnson2010quantum, SinglePhoton-Micro2014}. 
For the optical domain there are several  potential candidates:  quantum dots (QD) in coupled photonic crystal nano-cavities \cite{Coupled-CavityPRL2007}, nanofibers coupled to cavities with solid-state emitters (like NV centers in diamond, QD, rare-earth ions doped crystals) \cite{HakutaPRL2014} or natural atoms \cite{Thompson07062013, Shomroni22082014}. 
Open optical cavities \cite{JobezNJP2014} can also be used.        

However, it is worth noting that the qRAM operation requires the further improvement  of the control atom integration in the QED cavity. 
Various developing protocols of photon routing \cite{PhotonroutingPRA2011} can be applied for the effective photon addressing. 
The passive routing \cite{CQED-Gate-PRA2010, *CQEDGatePRA2011,  Shomroni22082014}, which does not require any control fields, seems especially promising.

\section{Conclusion} 
We proposed the scheme of qRAM  based on the single multi-qubit QM.  
The qRAM contains two coupled QED cavities with the control three-level atom and the resonant atomic ensemble. 
The multimodality is achieved by combining the time-domain control of the single three-level atom in the cavity together with the  photon echo multimode QM.  
The multi-time-bin photon state  establish the  addressing by setting the control atom in the superposition of transfer and blockade regimes. 
We found a series of impedance matching conditions for the physical parameters of the scheme providing the broadband efficient implementation of qRAM processes.
The performed analysis can be easily  extended to various variants of the photon echo QM protocols in the impedance matching QED cavities that essentially enlarges the potential for the experimental implementation.  
The proposed qRAM can be realized with current technologies of the circuit and cavity QED that seems promising for using in superconducting quantum computing and optical quantum communications. 
Unlike the bucket-brigade architecture \cite{QRAMPRL2008, *QRAM-Lloyd-PRA-2008, qRAM-China}, the proposed qRAM avoids the complex control on many optically coupled QM cells since it can be implemented in a single compact device containing multi-qubit quantum memory cell that seems promising for practice. 
Finally, we note that the proposed qRAM can be developed for off-resonant Raman atomic transitions that will provide the direct quantum storage on the long-lived atomic transition and facilitate the experimental implementation of the qRAM impedance matching conditions.    

\section{Acknowledgments}

The authors are grateful  to  A. V. Akimov, E. Giacobino, A. Lvovsky and L. Maccone for useful and stimulating discussions.  We thank the Russian Scientiﬁc Fund through the grant no. 14-12-01333 for financial support of this work. 
\bibliography{QRAM}
\onecolumngrid
\appendix*
\section{Supplementary materials}
\subsection{Hamiltonian and basic equations}
The Hamiltonian of the analyzed system is
$\hat{H}  = \hat{H}_0 + \hat{H}_1 $, where
\bea{}
& \hat{H}_0 &= \hbar \omega_0 \big( \sum_{j=1}^{N} S^{j}_z + S^{0}_z + a_{1}^{\dagger} a_{1} +a_{2}^{\dagger} a_{2} 
\nonumber\\ 
&+& \int d\omega \left( \sum_{m} \hat{c}^{\dagger}_{m}(\omega) \hat{c}_{m}(\omega) + b^{\dagger} (\omega) b(\omega)  \right)\big),  
\eea{}
is the basic Hamiltonian, and the perturbation part is
\bea{}
& \hat{H}_1  &= \hbar \sum^{N} \Delta_{j} \hat{S}^{j}_{z} + \hbar \delta S^{0}_z  \nonumber \\
&+& \hbar \int d\nu \nu \hat{b}^{\dagger} (\omega_0+\nu) \hat{b}(\omega_0+\nu) \nonumber \\
&+& \hbar \sum_{m}\int d\nu \nu \hat{c}^{\dagger}_{m} (\omega_0+\nu) \hat{c}_{m} (\omega_0+\nu) + \nonumber \\
&+&  \hbar f_1 \int d \nu (\hat{a}^{\dagger}_{1}\hat{b} (\omega_0+\nu)  +  H.C. ) \nonumber \\
&+& \hbar\left(  g_1 \hat{a}^{\dagger}_1\hat{S}^{0}_{-}  +f_2 \hat{a}^{\dagger}_1\hat{a}_2 + g_2 \sum_{j=1}^{N}\hat{a}^{\dagger}_2 \hat{S}^{j}_{-}+H.C.\right) \nonumber \\
&+& \hbar\sqrt{\frac{\gamma}{2\pi}} \sum_{m}\int d\nu   \left( \hat{c}_{m}(\omega_0 + \nu) \hat{S}^{0}_{+} + H.C. \right),
\eea{}

\noindent
where the first three terms are determined by  frequency detunings of the j-th atom $\Delta_j$ in QM, $\delta$ of the control atom and the detuning $\nu$ of the free field modes; 
four further terms are the interactions between the second cavity mode and atoms (with the coupling constant $g_2$), between free field modes and the first cavity mode (with the coupling constant $f_1$), and the interaction between the cavity mode and the control atom (with the coupling constant $g_1$), and the interaction  between the coupled cavity modes (with the coupling constant $f_2$); $\hat{S}_z^0$ 
$\hat{S}_z^j$ are the z projection of the spin 1/2 operators,
$\hat{S}^j_+$, $\hat{S}^j_-$ and $\hat{S}^0_+$, $\hat{S}^0_-$  are the transition spin operators of $j$-th and control atoms;
$\hat{a}^{\dagger}_{1,2}$ and $\hat{a}_{1,2}$ are arising
and decreasing operators of the $1$-st and $2$-nd cavity field modes;
$b^{\dagger}(\omega),b(\omega)$ are the bozonic operators of free propagating modes (
$\left[ b(\omega),b^{\dagger}(\omega')\right]= \delta (\omega-\omega')$). 

By assuming that the QED cavity modes and all the atoms and bath modes $c_m(\omega)$ are in the ground state, the initial wave function in the case of the input single photon field 
$\ket{\Psi(t \rightarrow - \infty)}=\int d\omega \alpha^o_{\omega} b^{\dagger}(\omega)\ket{0}$ and total wavefunction is given by Eq.(\ref{WF}). By using the well-known input-output formalism we obtain the following system of equations

\bea{}
\frac{d\alpha_1}{dt} =
 - i g_1 \beta_c -if_2\alpha_2- \frac{\kappa}{2} \alpha_1 + \sqrt{\kappa} \alpha_{in}(t) \label{cavity1},
\eea{}

\bea{}
\frac{d\beta_c}{dt} =
-i (\delta-i\gamma/2) \beta_c - i g_1  \alpha_1 \label{atom},
\eea{}

\bea{}
\frac{d \alpha_2}{dt} = -i g_2 \sum_{j=1}^{N} \beta_{j} -i f_2 \alpha_{1},  \label{cavity2}б
\label{alpha_2}
\eea{}

\bea{}
\frac{d\beta_j}{dt} = -i(\Delta_{j}-i/T_2)\beta_{j} -i g_2 \alpha_2,  \label{memory}
\eea{}

\noindent
where $\kappa = 2\pi f^2_1$, also we have phenomenologically added the weak decay constant $1/T_2$ for the atomic coherence QM caused by the interaction  with local fluctuating fields  
\bea{}\sqrt{\kappa} \alpha_{in}(t)= -if_1\int d\nu\alpha_{\nu}^o
e^{-i\nu t}. 
\eea{}

\noindent
Integrating equation (\ref{atom})
\bea{}
\beta_j(\tau) &=& -i g_2\int \limits_{-\infty}^{\tau} dt' \alpha_2 (t') {e^{-i(\Delta_j-i/T_2)(\tau-t')}}
_{\mid_{\lim_{\tau\gg \delta t} }} \nonumber \\
&\cong& -i 2\pi g_2\tilde{\alpha}_2(\Delta_j) e^{-i(\Delta_j-i/T_2)\tau} \label{atoms},
\eea{}

\noindent
where pulse duration of the light field $\delta t $ is assumed to be short enough in comparison with the atomic decoherence time of the atomic QM $\delta t \ll T_2$.
Inserting it into Eq.(\ref{alpha_2}) and using the  Fourier transform  
$\alpha_{1,2,in}(t) =\int d\nu
\tilde{\alpha}_{1,2,in}(\nu) 
\exp\{-i\nu t\}$,
and 
$\beta_c(t) =\int d\nu
\tilde{\beta}_c(\nu) 
\exp\{-i\nu t\}$ 
we find the following solution for the amplitudes of the control atom and two cavity modes:

\bea{}
\tilde{\beta}_c =  g_1 \frac{\tilde{\alpha}_1}{ (\nu-\delta+i\gamma/2)}, \\ 
\tilde{\alpha}_2 (\nu)= A_{2,1}(\nu)   \tilde{\alpha}_1 (\nu), \\
\tilde{\alpha}_1 (\nu) = A_{1,in}(\nu) \tilde{\alpha}_{in} (\nu),
\eea{}

where

\noindent
\bea{}
A_{1,in}(\nu) &=& \frac{i\sqrt{\kappa}}{\nu+ i\kappa/2  -  \frac{g^{2}_1}{ (\nu-\delta+i\gamma/2) } -  \frac{f^{2}_2}{\nu +i Ng^{2}_2 \tilde{G}(\nu) }},\nonumber\\
\\
A_{2,1}(\nu)&=&\frac{f_2}{\nu +i Ng^{2}_2 
\tilde{G}(\nu) }.
\eea{}

\subsection{Readout stage}
 The system of equations is almost the same as (\ref{cavity1}, \ref{atom}, \ref{cavity2},\ref{memory}), however, we take into account the absence of the driving field, inverted inhomogeneous broadening  at the reference time $t=\tau$:

\bea{}
\frac{d\alpha_1}{d t} =
 - i g_1 \beta_1 -if_2\alpha_2- \frac{\kappa}{2} \alpha_1, \\
\frac{d\beta_c}{d t} =
-i (\delta-i\gamma/2) \beta_c - i g_1  \alpha_1, \\
\frac{d \alpha_2}{d t} = -i g_2 \sum_{j=1}^{N} \beta_{j} -i f_2 \alpha_{1},  \\
\frac{d\beta_j}{d t} = i(\Delta_{j}+i/T_2)\beta_{j} -i g_2 \alpha_2. 
\eea{}
By taking into account the initial condition at $t =\tau$ in accordance with Eq.(2)
\beq{}
\beta_j(-\Delta_j,\tau) = i \sqrt{2\pi\kappa} \frac{g_2}{f_2}  F_S(\Delta_j) \alpha_{\Delta_j}^{0} e^{-i(\Delta_j-i/T_2) \tau}
\label{mapping},
\eeq{} 
\noindent 
and $\alpha_1=\alpha_2=\beta_c=0$, 
after the storage stage. 
By solving the linear system of equations (A.36)-(A.39), we find the following  spectral component of the irradiated single photon field 

\bea{}
\alpha_{\nu}(t)&=&
i \sqrt{2\pi\kappa} \frac{g_2}{f_2}
F_R(\nu)\beta(\nu,\tau)\cdot \nonumber\\
&& G_L(\nu) \exp\{-i\nu(t-\tau)-\tau/T_2\},
\eea{}

\noindent
where $F_R(\nu)$ indicates the spectral transfer function for the readout stage.  


\subsection{Blockade: wave function of atomic system}

For the blockade regime we find the following  Fourier image for the amplitude of the QM atomic excitation in the calculations Eqs.(A.36-A.39) for the retrieval stage:

\bea{}
&\beta&_B(\Delta,\nu)=
\frac{1}{1/T_2-i\nu-i\Delta}\cdot\nonumber\\ 
&&\left( \beta(\Delta,0) +
 \Delta_{in} J(\nu) (F_B(\nu)-1)  \int d\Delta' \frac{G(\Delta') \beta(\Delta',0)}{-i\nu+i\Delta'} \right),\nonumber\\
\eea{}
where the index "B" means the blockade regime and

\bea{}
J(\nu)=\frac{\kappa  (\kappa -2 i \nu )}
{\kappa ^2-4 i \kappa  \nu -8 \nu ^2}.\nonumber\\
\eea{}


In the case of the strong atomic blockade $2+4C\gg 1$:

\bea{}
&\beta&_B(\Delta,\nu)\cong
\frac{1}{1/T_2-i\nu-i\Delta}\cdot\nonumber\\ 
&&\left(\beta_0 (\Delta, 0) e^{-i\Delta \tau}
-  \frac{2\Delta^{2}_{in}J(\nu)}{ (\nu^{2}+\Delta^{2}_{in})}  \beta_0(-\nu,0) e^{i\nu\tau-\tau/T_2}\right),\nonumber\\
\eea{}
\noindent
where
 $\beta_0(-\nu,0)=i\sqrt{2\pi\kappa}\frac{g_2}{f_2}  F(-\nu) \alpha_{-\nu}^0 e^{-\tau/T_2}$.
By taking into account $J(\nu \approx 0)\approx 1$ and $\tau\ll T_2$, we obtain quite perfect recovery of the atomic amplitude for $|\nu|<\kappa$:

\bea{}
\beta_B(\Delta,|\nu|<\kappa)\mid_{\tau  \ll T_2}=
-\frac{1}{1/T_2-i\nu-i\Delta} \beta(\Delta,0).\nonumber\\
\eea{}
As it is seen in Eq.(A45), atomic dynamics during the irradiation and subsequent reabsorption of the echo pulse  leads  to the additional $\pi$-phase shift of the atomic coherence for $t\gg t_{echo}= 2\tau$.


\end{document}